# Towards a European Stratospheric Balloon Observatory – The ESBO Design Study

Philipp Maier*[a], Jürgen Wolf[a], Thomas Keilig[a], Alfred Krabbe[a], Rene Duffard[d], Jose-Luis Ortiz[d], Sabine Klinkner[a], Michael Lengowski[a], Thomas Müller[e], Christian Lockowandt[b], Christian Krockstedt[b], Norbert Kappelmann[c], Beate Stelzer[c], Klaus Werner[c], Stephan Geier[c,f], Christoph Kalkuhl[c], Thomas Rauch[c], Thomas Schanz[c]. Jürgen Barnstedt[c], Lauro Conti[c], Lars Hanke[c]

[a]Institute of Space Systems, University of Stuttgart, Pfaffenwaldring 29, 70569 Stuttgart, Germany; [b]Swedish Space Corporation, Torggatan 15, 17154 Solna, Sweden, [c]Institut für Astronomie und Astrophysik, Universität Tübingen, Sand 1, 72076 Tübingen, Germany, [d]Instituto de Astrofísica de Andalucía (CSIC), Glorieta de la Astronomía S/N, 18008 Granada, Spain, [e]Max-Planck-Institut für extraterrestrische Physik, Giessenbachstraße, 85741 Garching, Germany, [f]Institut für Physik und Astronomie, Universität Potsdam, Karl-Liebknecht-Str.24/25, 14467 Potsdam, Germany

## ABSTRACT

This paper presents the concept of a community-accessible stratospheric balloon-based observatory that is currently under preparation by a consortium of European research institutes and industry.

The planned European Stratospheric Balloon Observatory (ESBO) aims at complementing the current landscape of scientific ballooning activities by providing a service-centered infrastructure tailored towards broad astronomical use. In particular, the concept focuses on reusable platforms with exchangeable instruments and telescopes performing regular flights and an operations concept that provides researchers with options to test and operate own instruments, but later on also a proposal-based access to observations. It thereby aims at providing a complement to ground-, space-based, and airborne observatories in terms of access to wavelength regimes – particularly the ultraviolet (UV) and far infrared (FIR) regimes –, spatial resolution capability, and photometric stability. Within the currently ongoing ESBO *Design Study* (ESBO *DS*), financed within the European Union's Horizon 2020 Programme, a prototype platform carrying a 0.5-m telescope for UV and visible light observations is being built and concepts for larger following platforms, leading up to a next-generation FIR telescope are being studied. A flight of the UV/visible prototype platform is currently foreseen for 2021.

We present the technical motivation, science case, instrumentation, and a two-stage image stabilization approach of the 0.5-m UV/visible platform. In addition, we briefly describe the novel mid-sized stabilized balloon gondola under design to carry telescopes in the 0.5 to 0.6 m range as well as the currently considered flight option for this platform.

Secondly, we outline the scientific and technical motivation for a large balloon-based FIR telescope and the ESBO *DS* approach towards such an infrastructure.

## 1. INTRODUCTION

The idea of using stratospheric balloons to overcome the obstructions of Earth's atmosphere for astronomical observations is not new. Historically, the advantages were obvious: spacecraft did not exist and capabilities of planes were limited, leaving balloons as the only option to move instruments above most of the atmosphere. In current times, the benefits do not seem as clear: both spacecraft and planes provide powerful observation platforms and ground-based telescopes invest large efforts into compensating atmospheric influences. However, even in the era of nano- and microsatellites, space observatories are intrinsically expensive and bear operational limitations: development times are long, updates or corrections of the instrumentation are usually not possible after launch, operating material such as cryogenic coolant fluids

*pmaier@irs.uni-stuttgart.de; irs.uni-stuttgart.de

(see the Herschel Space Observatory) cannot be refilled or replaced. Furthermore, comparably conservative approaches towards new technologies are used to minimize risks of expensive failure. Ground-based and airborne telescopes, on the other hand, still suffer from some fundamental limitations imposed by the atmosphere. Among those are the inability to access certain spectral regions, including the UV and, from the ground, large parts of the infrared (IR), but also limitations to photometric accuracy due to atmospheric variations. The answers to many fundamental, yet still unresolved astrophysical questions, such as those about the detailed mechanisms of astronomical engines, the secrets of exoplanet atmospheres, or the distribution of water in our own solar system, closely linked with questions about its own formation and evolution, thus lay obscured behind this atmospheric curtain.

Particularly to access the UV and IR spectral regions, to obtain photometric stability, and to overcome the effects of turbulence, though, it is sufficient to move into the high stratosphere at 30 to 40 km altitude, above 99% of Earth's atmospheric mass [1]. Balloon-based observatories can cater this region while maintaining accessibility and flexibility of instruments similar to ground-based observatories.

Nevertheless, the use of balloon-based telescopes is limited. With some notable exceptions [1], [2], modern balloon-based telescopes have mostly served a single purpose and did not fly more than a few times. It seems likely that the main reasons are a combination of the challenges associated with large balloon payloads (such as safe payload recovery and precise pointing) and the specialized expertise required for large stratospheric balloon missions that most astronomical research groups do not have. A visible trend towards providing more specialized equipment to address the needs of balloon-based astronomical telescopes exist, such as the Wallops Arc Second Pointer [3] or the CNES Generic Architecture for Multi-Mission Pointing Gondolas [4]. A true operating institution that helps to overcome the abovementioned and related challenges by providing observing time and instrument space on observatories, as existing in the ground- and space-based domains (such as the European Southern Observatory (ESO) or the European Space Agency (ESA), respectively), however, is missing for balloon-based astronomical observations.

The goal of ESBO is to form such an institution in Europe and ESBO *DS* shall pave the way towards it.

This paper is consequently structured into two parts. After a general outline of the concept and the objectives of both ESBO and ESBO *DS* in section 2, sections 3 to 7 describe the technical developments of the prototype platform STUDIO (Stratospheric Ultraviolet Demonstrator of an Imaging Observatory) that are currently under way within ESBO *DS*. Sections 8 and 9 outline the motivation for the future development towards a FIR observatory and provide an outlook onto the potential evolution of ESBO in the future.

## 2. ORISON, ESBO, AND ESBO *DS*

### 2.1 Findings of the precursor project ORISON

Over the last two years, the H2020-funded project ORISON (innOvative Research Infrastructure based on Stratospheric balloONs) assessed interests and scientific needs within the (mostly European) astronomical community with regard to balloon-based research infrastructures and studied the general feasibility of a balloon-based observatory (http://www.orison.eu). ORISON showed that extended interest in stratospheric observation conditions exists within the European astronomical community and that the concept of versatile, mid-sized balloon telescopes operated as an observatory is feasible. The work also showed, however, that further technological development is required and that significant potential lies with larger flight infrastructures than originally foreseen under ORISON, particularly for the near and far infrared spectral ranges [5].

The plans for ESBO as well as the ESBO *DS* project pick up from these conclusions.

### 2.2 ESBO

ESBO represents the idea of the larger observatory, of a service provider that offers a wide scientific community access to the astronomical observation conditions in the high stratosphere – both in terms of instrument space and observing time. The concept for making this idea feasible and competitive includes the following key considerations:

- The observatory will offer regular flights with reusable hardware to ensure efficient and fast turnaround times in between flights. To ensure the reusability, safe recovery of the payload will need to be guaranteed. Safe recovery and re-flight will also allow the amortization of hardware costs over a number of flights.

- The observatory will strive to fully use current ballooning technology to offer longer flights in the order of weeks to months.
- To use the infrastructure efficiently and to open it to diverse scientific applications, instruments and/or telescopes on the flight platforms will be exchangeable. This may include that several flight platforms are ready for use at the same time.
- The regular touchdown every several months or weeks in combination with the possibility to exchange instruments will allow the refilling of consumables (especially cryogens) as well as easy and fast profit from technology developments and instrument upgrades.
- To provide access for a wide community, both instrument flight opportunities and open observation time access (either via shared time on principal investigator (PI) instruments or via facility instruments) will be foreseen.
- The mission of the observatory will be that of a service provider – both to instrument builders and observers.

These considerations require that the ESBO infrastructure does not only include flight systems (FS), but that it also offers tools and procedures for the instrument preparation, proposal, observation, and data reduction phases. Figure 1 illustrates the full envisioned ESBO infrastructure.

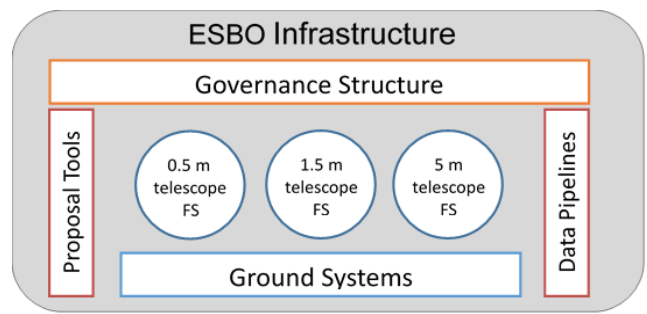

Fig. 1: Illustration of ESBO infrastructure elements (FS = flight system)

The implementation of ESBO is planned as a step-by-step approach, building up both flight systems and the operating institution. In the short term, a small flight system with a 0.5 m aperture telescope tailored for the UV and visible range is foreseen. It will potentially be followed by a mid-sized (1.5 m aperture class) flight system tailored for the NIR, eventually leading to a FIR flight system likely in the 5 m aperture range in the long term.

## 2.3 ESBO *DS*

The ongoing ESBO *Design Study* represents, after ORISON, the second step towards ESBO. Under ESBO *DS*, the full infrastructure is being conceptually designed and a prototype UV/visible flight system is being developed and built. Particular emphasis thereby lays on technological development with regard to safe recovery of payloads, modular and scalable gondolas, and versatile and precise pointing/image stabilization. The activities within ESBO *DS* can be summarized as follows:

- Conceptual design of the technical aspects (including flight systems) of ESBO;
- Conceptualization of an operations and governance framework within which ESBO can be maintained and expanded in the long term;
- Development of a roadmap to full capabilities, particularly for FIR observations, including the identification of critical required developments and associated costs;
- Demonstration of the maturity of the critical technologies for the first step (UV/visible platform) via the STUDIO prototype platform;
- Demonstration of a UV instrument and visible light photometric measurements on the STUDIO platform;
- Provision of the STUDIO prototype platform for scientific use and exploitation after the ESBO DS project.

## 3. TECHNICAL MOTIVATION FOR THE UV/VISIBLE PLATFORM (STUDIO)

The prototype UV/visible platform STUDIO currently under development will serve two purposes: on the one hand, it will serve as a technology demonstrator and testbed for critical technologies required for the further implementation of ESBO. On the other hand, the platform will also function as a testbed for a new microchannel plate (MCP) detector for the UV and will be available for first scientific test flights after the end of the ESBO *DS* project.

The critical technologies to be demonstrated include a modular and scalable gondola to accommodate different astronomical payloads and their respective needs, systems for safe recovery, and a versatile, highly precise image stabilization system. The modular and scalable gondola is described in further detail in section 6. For safe recovery, the foreseen systems include autonomous steered parafoils for controlled landings, in combination with a suitable gondola design, as well as a shock damping mechanism to decrease the parachute opening load. The image stabilization system will be a two-stage system, with coarse pointing control provided by the gondola, further described in section 6, and a fine image stabilization system included within the optical system described in further detail in section 5.

The chosen image stabilization approach motivates a visible light camera in addition to the main scientific UV instrument to be used as a focal plane sensor for the image stabilization system, but also usable as an add-on scientific instrument to e.g. measure photometric stability and to complement the UV scientific observations.

In the following sections, the STUDIO will be described in more detail. Section 4 will cover the scientific motivation, section 5 the scientific and technology demonstration payload, section 6 the balloon gondola, and section 7 the foreseen flight option.

## 4. UV SCIENTIFIC MOTIVATION

Astronomical observations in the UV at wavelengths below ~320 nm are not possible from the ground because of atmospheric extinction by ozone. However, at a height of 40 km, observations are feasible down to ~200 nm. Only in the range 240-260 nm, residual ozone reduces the transmission to ~20%. Therefore, the balloon-borne STUDIO platform enables UV observations that would otherwise only be possible with space-based telescopes.

STUDIO comprises a 50 cm aperture telescope and two simultaneously operating imaging instruments as a UV and a visible/NIR channel. The main scientific instrument will be the UV instrument covering the 180-330 nm band at an angular resolution of ca. 1.03 arcsec and with a field of view of 30 arcmin x 30 arcmin. Two science cases motivate the UV scientific part of STUDIO, which are shortly described in the following.

### 4.1 Search for variable hot compact stars

Hot and compact stars are the rather short-lived end stages of stellar evolution. They comprise the hottest white dwarfs (WDs) and hot subdwarfs. A significant fraction of them show light variations with periods ranging from seconds to hours. Among them are diverse types of pulsators, which are important to improve asteroseismic models. Others are members of ultracompact binaries (e.g., WD+WD pairs) and are strong sources of gravitational wave radiation and crucial calibrators for the future space mission eLISA. They are also regarded as good candidates for the progenitors of thermonuclear supernovae. Furthermore, compact binaries are formed via common envelope evolution and are important to study this poorly understood phase of binary evolution.

Hot compact stars have so far been studied predominantly at high Galactic latitudes. Due to their very blue colors they stick out in old stellar populations like the Galactic halo. However, the density of stars at high Galactic latitudes is rather small and those objects therefore very rare. Due to the 1000-times higher stellar density, the Galactic disc should contain many more of those objects. Searches in the Galactic plane are desirable but the identification of these faint stars is hampered by the dense, crowded fields. But not so in the UV band. The hot stars are much easier to detect there, because their emitted flux is increasing towards the UV, while the flux of the majority of other stars decreases because of their lower temperatures. Surveying the Galactic plane with a UV imaging telescope will uncover many new variable hot stars.

### 4.2 Detection of flares from cool dwarf stars

Red dwarf stars (spectral type M) are hydrogen-burning main sequence stars like our Sun, but less massive, cooler and less luminous. The large majority of the stars in our Milky Way belongs to this group. Red dwarfs emit most of their radiation in the visible and near-infrared wavelength regions. Their UV and X-ray emission, despite being energetically a minor contribution to the overall radiation budget, ionizes material surrounding the stars and is, therefore, of central interest for the evolution of planets and other circumstellar matter. This high-energy emission of red dwarf stars is highly dynamic.

One characteristic phenomenon are flares that are stochastic brightness outbursts resulting from reconfigurations of the stellar magnetic field. During such flares, these normally faint stars become much brighter for the duration of minutes. A strong emission line of ionized magnesium (Mg II) at 280 nm, covered by the STUDIO instrument, can carry up to 50% of the near-UV flux during flares. Up to now, no systematic monitoring of "flare stars" exists. Consequently, the flare occurrence rate is unknown as well as the flare energy number distribution. Particularly interesting for the study of the physics of flares is their multi-wavelength behavior (time lags, relative energy in different bands). However, only a few simultaneous UV and optical observations exist. STUDIO enables such observations by continued monitoring (over hours or multi-epoch) of stars across the field or by focusing on prominent objects.

# 5. STUDIO PAYLOAD

## 5.1 UV Instrumentation

The STUDIO payload comprises a 50 cm telescope with a beam splitter for a visible and an ultraviolet channel. The main science instrument is an imaging and photon counting microchannel plate (MCP) UV detector, developed and built by the "Institut für Astronomie und Astrophysik Tübingen" (IAAT). It is a successor to the Echelle detector whose development and flights for the ORFEUS missions are part of the space heritage of the IAAT [6].

### 5.1.1 Working principle

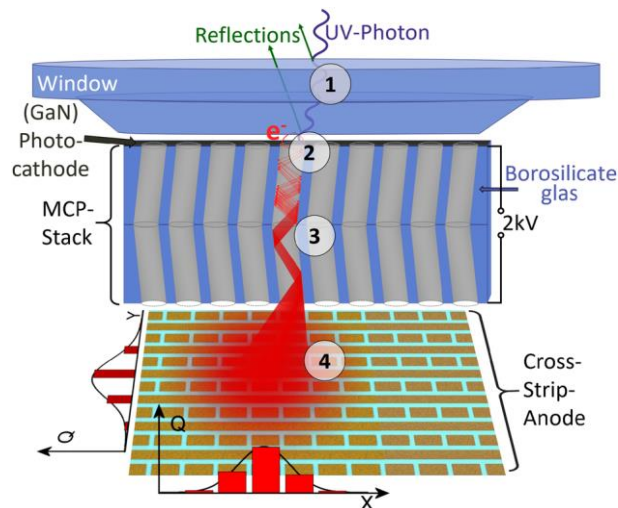

Fig. 2: Working principle of the UV detector (taken from [7]).

A detailed view of the working principle and the individual components can be found in [7]. In the following, a brief description of the enumerated parts in figure 2 is given:

1. The detector is sealed by a UV-transparent *window* (fused silica/MgF$_2$/LiF) to maintain an ultra-high vacuum, which is needed for a proper photocathode operation.

2. The *photocathode* converts a photon into a photoelectron with a conversion rate depending on the photocathode material and the wavelength of the incoming photon. A quantum-efficiency of up to 70% at 230 nm seems achievable for GaN [8].

3. In the *microchannel plates* (MCPs) a high voltage of about 2 kV accelerates the incident photoelectron, resulting in a charge cloud (about $10^5$ electrons) at the bottom side of the MCP stack. The position information of the incident photoelectron is preserved in the MCPs as the center of charge of the electron cloud.

4. The electron cloud is accelerated towards a *cross strip anode* (64 strips in X direction and 64 strips in Y direction) and places a certain amount of charge on several strips of the anode.

The charge signals of the 128 anode-strips are amplified by the BEETLE chip (for details of the BEETLE chip see [9]), which was developed by the "Max Planck Institut für Kernphysik" in Heidelberg for the LHCb experiment. The output of the BEETLE chip is converted into digital signals and sent to the processing unit, a Virtex-5 FPGA. The implemented centroiding algorithm will calculate the center of charge for each event with an accuracy of 1/32 of the distance between two anode strips. This results in an image size of about 2000 x 2000 pixels.

**5.1.2 Overview of the instrument on STUDIO**

The UV detector will cover the wavelength band from 180 nm to 330 nm. The detector has a sensitive area of 39 mm diam. With an image size of 2048 × 2048 pixels the angular resolution will be 1.03 arcsec per pixel. The two possible operation modes are the image integration mode and a photon-by-photon readout mode, which would allow for an offline correction due to a possible pointing jitter of the telescope. The detector will be capable of processing about 200 000 – 300 000 detected photons per second. It is also foreseen to implement a filter wheel carrying two filters: Sloan u ($\lambda_{mean} = $ 349.8 nm, [10]) and Galex NUV ($\lambda_{mean} = 231.57$ nm, [11]) as well as an open position. The mass of the whole detector is about 4.4 kg (including high voltage power supply and cables). The power consumption still needs to be determined but will be below 19 W (peak).

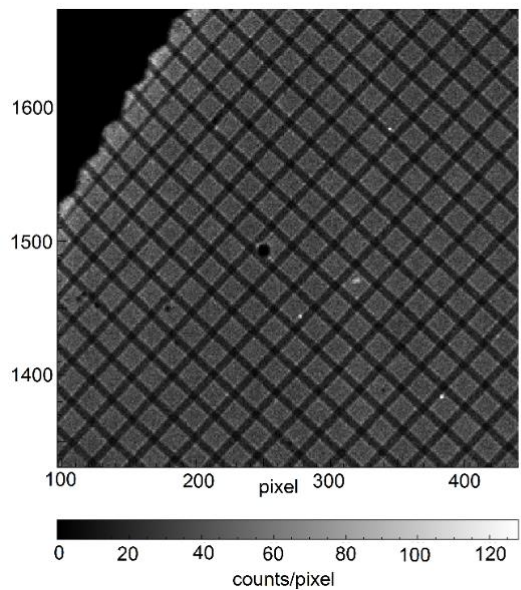

Fig. 3: Detector image of a test lattice for demonstration of the centroiding algorithm. The lattice constant is 0.4 mm and the lattice was placed about 1 cm in front of the detector. Light source was a mercury lamp with its main emission line at 254 nm. On the upper left corner one can see the edge of the MCPs. One can also observe some small hot spots as well as a bigger black spot where the MCPs might be defect.

**5.2 Optical layout**

The baseline optical system of the STUDIO payload is a telescope with a primary mirror of 500 mm aperture in a Cassegrain configuration. A focal range of N = 8, i.e. a focal length of f = 4 m thereby leads to an acceptable length of the optical tube assembly for integration into the balloon gondola of about 1.6 m. For the detector configuration with the main

UV instrument and a visible light tracking camera as shown in table 1, first performance estimates have furthermore shown that a classical Cassegrain configuration will likely be preferable over a Ritchey-Chretien configuration. While a Ritchey-Chretien configuration would bring the visible channel closer to being diffraction limited, the sampling of the foreseen camera would not allow for sampling at the diffraction limit. On the other hand, the performance of a Ritchey-Chretien is significantly less favorable at the edges of the large UV sensor than that of the classical Cassegrain.

The telescope instruments platform (TIP) attached to the back of the optical tube assembly will house the beamsplitter, the two instruments, as well as a filter wheel for each instrument.

Table 1. Basic geometrical parameters of STUDIO detectors

| Detector Geometries | | | |
|---|---|---|---|
| UV Detector | | Baseline Visible Camera | |
| Detector Size | 39 mm diam. | Detector Size | 13.3 mm x 13.3 mm |
| Pixel Size | 19 µm x 19 µm | Pixel Size | 13 µm x 13 µm |
| Pixel Array | 2048 px diam. | Pixel Array | 1024 x 1024 |
| Pixel Scale | 1.03 arcsec/px | Pixel Scale | 0.67 arcsec/px |
| Field of View | 35.2 arcmin diam. | Field of View | 11.4 x 11.4 arcmin$^2$ |

### 5.3 Image stabilization system

The scientific UV instrumentation requires a pointing stability of 0.5 arcsec. Reaching such precise stability is only achievable with a multi-stage system. STUDIO will employ a coarse attitude control system for elevation and azimuth stabilization down to ca. +/- 40 arcsec (see further information in the following section) and a fine image stabilization system within the optical system for the remaining jitter compensation. The image stabilization system will employ a closed-loop control system with a fast steering mirror and the visible camera as the guiding sensor.

## 6. BALLOON GONDOLA

### 6.1 Modular and scalable gondola

The balloon gondola will be housing the telescope or scientific payload and all required support systems for the payload. Balloon gondolas are normally developed for a specific mission and payloads are usually flown a couple of times at the most, depending on the scientific goals program. However, the ESBO gondola will be designed to be re-flown many times. The support systems will be developed with high modularity, scalability, flexibility and reusability as a baseline. This will enable easy adaption of the gondola to different telescopes and missions. The reusability aspect will make it possible to fly the instrument frequently and maximize the flight and observation time with the different telescopes and at the same time reduce cost and time for refurbishment of gondola and support systems in between flights. Some aspects that are important for the gondola design are:

- Mechanical stiffness;
- Thermal protection for instruments;
- Easy integration and disintegration;
- Robustness.

The gondola will be equipped with support systems for the different telescopes/instruments to be flown such as:

- Mechanical gondola structure;
- Power system;
- Communication system;
- Thermal control system;
- Gondola control and coordination system;
- Landing and recovery system.

The subsystems will be reused from mission to mission in order to reduce cost and development time for the program. The focus can then be on the development of instruments, telescopes, and flight or observation time.

## 6.2 Landing

At termination of the flight, the gondola is separated from the balloon and traditionally descends on a non-steerable parachute down to the ground following the wind. Owing to the flight trajectories and the maximum use of observation time, the landing and recovery of the gondola in most cases take place in remote areas. Depending on the terrain and wind conditions, the landing can be rough, causing damage to both the gondola and instrument. To mitigate this and reduce the risk of damages, especially important in the ESBO case where the gondola and instrument are foreseen to re-fly often, novel landing systems will be foreseen such as steerable parafoils and potentially air bags. The use of steerable parafoils will not only allow the reduction of structural landing loads, but also a choice of the landing zone within the operating radius of the parafoil system, promising a significant reduction of recovery efforts.

## 6.3 Attitude control system

The telescope has to be pointed towards the desired object on the sky for observation in azimuth and elevation. As mentioned above, this will be achieved by a two-stage system. The first stage will be performed by a coarse control system, turning the complete gondola 360° for azimuth and tipping the telescope around its mounting axis for elevation pointing. The coarse system will function largely independently from the inner stage, employing its own attitude sensors, including differential GPS, gyroscopes, and a star tracker. There will be a limitation around zenith elevation due to the balloon obscuring the field of view. The limitations are dependent on the size of the balloon and the distance between the balloon and the gondola. The coarse pointing system to be developed and used for ESBO will be based on previous designs used on the astronomic PoGO balloon missions [12],[13].

# 7. FLIGHT OPTIONS

Large stratospheric balloons follow the winds. The means of controlling the flight and flight trajectory are twofold. The first is to use known and stable wind conditions. In some places on the globe the wind directions are very predictable and stable during specific periods of the year at the flight altitudes of 30-40 km. These conditions can be used to achieve a predictable flight trajectory. One example is the polar region; during the summer the wind at these altitudes is always in a westerly direction in the northern polar hemisphere. This makes it possible to fly from Esrange Space Center in northern Sweden to northern Canada or Alaska in approximately one week. The second method is to use wind layers with different wind directions and speeds at different altitudes. The altitude of the balloon can be adjusted by dropping ballast or releasing gas from the balloon.

The first flight of the ESBO flight prototype is foreseen in the 2021. The first flight will be performed from Esrange Space Center in northern Sweden, taking advantage of the seasonal "turnaround conditions" of the polar stratospheric wind systems that allow flights up to 40 h practically above Esrange.
The vision of the ESBO project, however, is to offer a balloon based stratospheric observation platform that can provide a maximum of observation time for the scientific community. This requires long and numerous flights with the ESBO platform. As the favorable wind and weather conditions for balloon flights change over the seasons globally, reaching maximum observation times requires to also change the launch and flight sites seasonally. The long-term plans for ESBO therefore foresee flights from different places on the globe, potentially with different launch site providers. Different launch sites are also necessary to cover observations over the complete hemisphere.

# 8. MOTIVATION FOR A FAR INFRARED TELESCOPE

## 8.1 FIR Technical background and current situation

As the UV region, the far infrared spectral region is equally inaccessible from the ground (see also figure 4). It does, however, become accessible at lower altitudes compared to the UV regime, so that the airborne Stratospheric Observatory for Infrared Astronomy (SOFIA) can observe in the FIR at 13 km flight altitude. Due to its fundamental inaccessibility from the ground, the FIR region cannot benefit from the technological improvements of large ground-based observatories such as atmospheric correction techniques or interferometry. FIR observations are thus solely reliant upon air-, balloon-, or space-borne capabilities. With the last large FIR space observatory, Herschel, having reached the end of its mission after running out of cryogenic coolant fluids, the FIR observatories currently available are SOFIA and sporadically flying balloon observatories (such as BLAST, STO, or PILOT [2],[14],[15]). For the next step of science following Herschel and SOFIA, improvements in two domains are necessary, as expressed by the scientific community: higher sensitivity and better angular resolution. The first need would be addressed by the Space Infrared Telescope for Cosmology and Astrophysics (SPICA, 2.5 m telescope diameter) [16], or, in combination with high spectral resolution, by the Far-Infrared Spectroscopic Explorer (FIRSPEX, 1.2 m telescope diameter) [17]. Both were proposed for the M5 mission of ESA's Cosmic Vision, with SPICA recently having been downselected as one of the three finalists (the M5 estimated launch date is in 2029/2030). Space telescope concepts with larger telescope sizes to address the need for angular resolution are being proposed as well, such as the Origins Space Telescope (OST) or the Thinned Aperture Light Collector (TALC, to be proposed for L4 of ESA's Cosmic Vision) [18]. Due to the logistical and technological challenges, however, their implementation would be far ahead in the future (the next L4 launch opportunity is not foreseen until 2042).

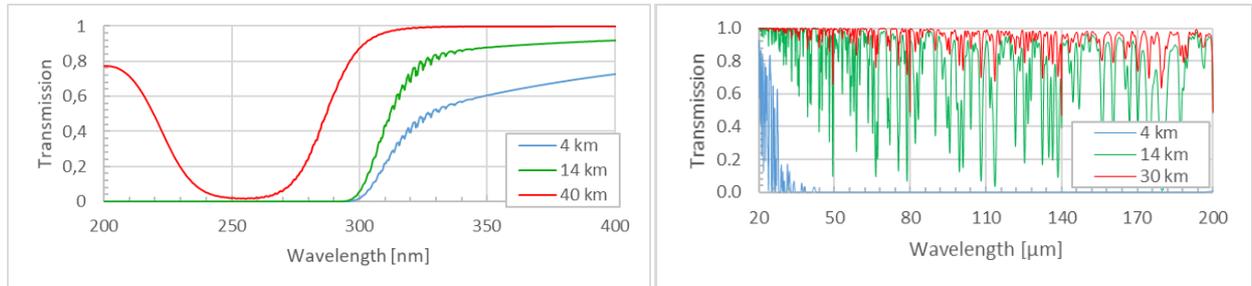

Fig. 4: Left: Atmospheric transmission from MODTRAN simulations at different altitudes in the UV. (Data at 40 km for a zenith angle of 30 deg, at 4 km and 14 km of 0 deg) Right: Atmospheric transmission from ATRAN [19] simulations at different altitudes in the far infrared. (4 km is roughly the altitude of the large ground-based observatories (ALMA, VLT), 14 km is roughly the SOFIA flight altitude).

The increase of sensitivity can likely best be addressed by space telescopes with cooled optics. Large mirrors, as required for high angular resolution, however, pose significant challenges to space missions which, in order to overcome the size limitation imposed by launcher fairings have to rely upon complex and expensive unfolding mechanisms such as used on the James Webb Space Telescope [20]. Similarly, the 2.7 m telescope on SOFIA is about the largest telescope size that can be accommodated on an existing plane that reaches the required altitude. Stratospheric balloons therefore offer an attractive platform for large aperture, high spatial resolution FIR telescopes.

## 8.2 Scientific motivation

Angular resolution scales with wavelength/telescope-diameter, which allows near and mid infrared missions at, e.g., 10 µm wavelength to achieve angular resolutions in the range of 2 to 3 arcsec with telescope diameters in the range of e.g. 0.85 m like the Spitzer Space Telescope. Achieving a similar angular resolution at 100 µm requires a telescope with proportionally larger diameters. While subarcsecond scales can now be achieved across many wavebands by ground-based single dish 8-10 m-class telescopes (ESO Very Large Telescope, Keck, Subaru) or by deploying interferometry (e.g. Atacama Large Millimeter/submillimetre Array (ALMA)), far infrared astronomy is still lagging behind because the largest telescope aperture sizes in this regime were Herschel with 3.5 m and is SOFIA with 2.5 m, providing angular resolutions between 10 and 7 arcsec at 100 µm wavelength.

The second argument for a large telescope diameter is the presence of interstellar cirrus. Fractal structures of dust and gas in the foreground dilute the radiation of a small astronomical target like a distant galaxy in a way that it becomes indistinguishable from the foreground, if it is smaller or comparable in diameter than the diffraction limited resolution of the telescope. This confusion limit, as it is called, is severely hampering the investigation not only of galactic but also of extragalactic targets and can only be overcome by larger apertures or interferometry. The confusion limit is less severe for spectroscopy due to the limited bandwidth and the presence of spectral features compared to photometric investigations. However, it remains an issue for all far-infrared investigations.

There is a trade-off between achieving a larger telescope diameter and controlling other parameters. For example, actively cooling the telescope will boost the sensitivity of the instrumentation on board, mostly due to lower background radiation level. Lower telescope temperatures can, e.g., be achieved by a higher flight altitude (SOFIA versus balloons). Increasing the efficiency of the observatory can also be achieved by increasing the number of detectors (multiplexing) or by adding observing time. The confusion limit, however, can only be overcome by a larger effective aperture, making the choice of the telescope diameter a fundamental one for the layout of an observatory.

Such trade can also be studied with Herschel vs. SPICA. While Herschel was a passively cooled telescope system with 3.5 m diameter, SPICA is a proposed actively cooled FIR Space telescope in the 2 m range. SPICA has sparked a discussion about the scientific value of diameter vs. cooling, since the 2.5 m diameter of the telescope is the same as SOFIA, which makes both of them very comparable in their confusion limits.

A long life-time balloon borne observatory, which touches ground every couple of weeks will much easier and faster profit from technology development, instrument upgrades, as well as increase multiplex advantages compared with a space mission, which generally does not allow for maintenance or upgrades. Therefore, for the far-infrared telescope the largest diameter feasible was chosen in order to be able to clearly go beyond the Herschel capabilities in particular with respect to the confusion limit.

ESBO *DS* therefore focuses on studying the feasibility of an effective telescope aperture of 5 m diameter and potentially elliptical shape, providing an angular resolution (1.22 $\lambda/D$) of 5.0 arcsec at 100 µm wavelength, comparable to the angular resolution achieved by Spitzer at 17 µm and Herschel at 70 µm wavelength.

The addition of this truly ground-breaking flight system is regarded as the long-term goal of the ESBO efforts.

## 9. ESBO IN THE FUTURE

### 9.1 Operational Concept

At the heart of the ESBO concept is the idea of creating a service provider – for instrument developers and for general observers. This concept is common among ground-based, airborne, and also some space-based telescopes. The telescope hardware and flight platform will be provided and operated by a central organization, or a consortium. Particularly in the short- and mid-term, ESBO will provide flight opportunities for instruments developed by scientific teams to be flown on the telescope ("PI Instruments"), within a shared-time approach. In the long term, turning instruments into facility instruments might be considered as an option.

While part of the available observation time will be reserved, the ESBO concept relies on making a share available to the community via open calls for proposals. A time allocation committee reviews the proposals, based on which observation time is granted.

The approach will be similar to the operating fashion of SOFIA or the ESO telescopes. Similar programs also exist for general-purpose balloon experiments, where a balloon gondola is provided by an operator and experiment proposals are invited, such as the French/Canadian "Stratos" program or the DLR/SSC/ESA BEXUS program.

ESBO will extend this approach to astronomical applications by offering: (i) Peer-reviewed, proposal-based access to stratospheric observations for the astronomical community; (ii) Regular and well controlled flight opportunities for astronomical instruments with secured recovery and return of instruments.

### 9.2 Timeline

The plans for ESBO foresee a step-wise development, which is outlined in figure 5. The current prototype development and conceptual design under ESBO *DS* will be concluded in 2021 and lead to the STUDIO prototype flight shortly thereafter. This flight will simultaneously serve as a first science precursor of the 0.5 m flight system. Further scientific and technology test re-flights of the modified STUDIO payload and gondola are foreseen thereafter. ESBO *DS* will also serve to develop a user group for further payloads, also for a mid-sized flight infrastructure to be potentially added in the 2023/2024 timeframe. Regular operation of the 5-m-class far infrared flight system is estimated in a 15 year timeframe.

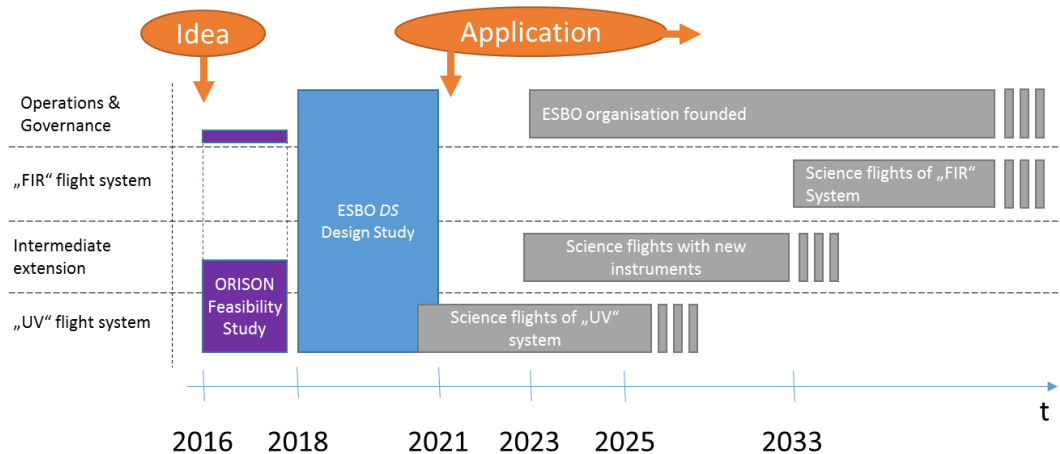

Fig. 5: Preliminary ESBO timeline. All future projections (grey) are preliminary plans based on current estimates and will be refined in a development roadmap during ESBO *DS*.

## ACKNOWLEDGEMENT

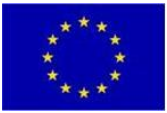
ESBO *DS* has received funding from the European Union's Horizon 2020 research and innovation programme under grant agreement No 777516.